\documentclass[11pt]{article}

\usepackage{epsf}
\usepackage{cite}
\usepackage{hyperref}

\def\1ad{\mbox{\normalsize $^1$}}
\def\2ad{\mbox{\normalsize $^2$}}
\def\3ad{\mbox{\normalsize $^3$}}
\def\4ad{\mbox{\normalsize $^4$}}
\def\5ad{\mbox{\normalsize $^5$}}
\def\6ad{\mbox{\normalsize $^6$}}
\def\7ad{\mbox{\normalsize $^7$}}
\def\8ad{\mbox{\normalsize $^8$}}

\def\beq{\begin{equation}}                     %
\def\eeq{\end{equation}}                       %
\def\bea{\begin{eqnarray}}                     
\def\eea{\end{eqnarray}}                       
\def\dj{\hbox{d\kern-0.347em \vrule width 0.3em height 1.252ex depth
-1.21ex \kern 0.051em}}

\def\half{{1\over 2}\,}

\def\Tr{{\rm Tr\,}}

\def\pt{\partial}

\def\CD{{\cal D}}

\def\shalf{{\mbox{$\half$}}}

\def\Dirac{\,\raise.15ex\hbox{/}\mkern-13.5mu D}
\def\dirac{\,\raise.15ex\hbox{/}\kern-.57em \partial}
\def\pslash{\,\raise.15ex\hbox{/}\kern-.57em p}

\begin{document}

\newcommand{\sheptitle}
{A note on the topological order of noncommutative Hall fluids}
\newcommand{\shepauthora}
{{\sc
 J.L.F.~Barb\'on  and D. Gerber}}

\newcommand{\shepaddressa}
{\sl
Instituto de F\'{\i}sica Te\'orica  IFT UAM/CSIC \\
 Facultad de Ciencias C-XVI \\
C.U. Cantoblanco, E-28049 Madrid, Spain\\
{\tt jose.barbon@uam.es} \hspace{0.5cm} { \tt daniel.gerber@estudiante.uam.es}}

\newcommand{\shepabstract}
{
We evaluate the ground state degeneracy of noncommutative
Chern--Simons models on the two-torus, a quantity that is
interpreted as the ``topological order" of associated phases of Hall
fluids. We define the noncommutative theory via T-duality from an
ordinary Chern--Simons model with non-abelian 't Hooft magnetic
fluxes. Motivated by this T-duality, we propose a discrete family of
noncommutative, non-abelian fluid models, arising as a natural
generalization of the standard noncommutative Chern--Simons
effective models. We compute the topological order for these
universality classes, and comment on their possible microscopic
interpretation.

}

\begin{titlepage}
\begin{flushright}

{IFT/2007-07-25\\ 
}

\end{flushright}
\vspace{0.5in}
\vspace{0.5in}
\begin{center}
{\large{\bf \sheptitle}}
\bigskip\bigskip \\ \shepauthora \\ \mbox{} \\ {\it \shepaddressa} \\
\vspace{0.2in}
\vspace{1in}

{\bf Abstract} \bigskip \end{center} \setcounter{page}{0}
 \shepabstract

\vspace{2in}
\begin{flushleft}
\today
\end{flushleft}


\end{titlepage}



\section{{Introduction}}

\noindent

\setcounter{equation}{0}

Hydrodynamical models of quantum Hall fluids contain a universal
Chern--Simons term
\begin{equation}\label{cse}
 S_{\rm eff} = {k\over 4\pi}
\int a\wedge da + \dots
\end{equation} that dominates the long-distance dynamics
\cite{ZhangHanssonKivelson:1988wy,Read:1988mp,GirvinMacDonald:1987fp,FrohlichZee:1991wb,FrohlichKerler:1990xz,Zee:1996fe}.
The description of the fluid density perturbations in terms of a
$2+1$ dimensional gauge field is characteristic of superfluids, the
$U(1)$ gauge symmetry being a remnant of the underlying quantum
permutation symmetry of identical particles forming the fluid
\cite{BahcallSusskind:1991an}. The emergence of a particular
Chern--Simons term with nonvanishing integral coupling, $k$, is
related to the broken parity and time reversal invariance by the
magnetic field and to the incompressibility of the fluid, so that
density perturbations have a gap proportional to $\sqrt{k}$. In the
simple phases such as the Laughlin series\cite{Laughlin:1983}, $k$
is an odd integer equal to $1/\nu$ with $\nu$ the filling fraction.
All the universal properties of those Hall fluids, such as the
spectrum of edge excitations, quasiparticle statistics and
topological order parameters, are controlled by the single integer
$k$. One particularly novel property of these fluids is a new kind
of vacuum order, the so-called {\it topological order}: the
structure of the Hilbert space is partially determined by the
spatial topology of the fluid. In terms of the effective theory
(\ref{cse}), the topological order is given by the finite dimension
of the Hilbert space: ${\cal D}_g \equiv {\rm dim}\,{\cal
H}(\Sigma_g) =k^g$ on a Riemann surface of genus $g$. For a general
review and references on the physics of the Hall effect see for
example \cite{Ezawa:2000ae}.

For more general Hall phases, there exist some standard
generalizations of (\ref{cse}), such as the multi-component abelian
fluid models (c.f.
\cite{WenZee:1992uk,BlokWen:1990an,BlokWen:1990mc,BlokWen:1990nd,LopezFradkin:1998ih,LopezFradkin:1991wy}),
\beq\label{csm}
 S_{\rm eff} = \sum_{I,J} {K_{IJ}
\over 4\pi} \int a_I \wedge da_J + \dots \eeq with $K_{IJ}$ an
appropriate matrix of integers that characterizes completely the
Hall fluid, including the spectrum and statistics of quasiparticles
and topological orders. For example, the topological order on a
genus-$g$ Riemann surface is given by ${\cal D}_g = |\det (K)|^g$.
The various components of the fluid could be rooted on specific
properties of the system, such as the case of multi-layered
electrons, or the relevance of spin degrees of freedom. In other
cases, they just provide a phenomenological parametrization of the
microscopic dynamics. Some of these models with fluid $U(1)^{\otimes
n}$ symmetry might admit a more accurate description in terms of
non-abelian Chern--Simons topological theories, leading to
non-abelian statistics of quasiparticles
\cite{MooreRead:1991ks,FradkinNayakTsvelikWilczek:1997ge}.

More recently, it was proposed in \cite{Susskind:2001fb} that a more
accurate version of the hydrodynamics, capable of keeping track of
the `granular' character of the electrons, is provided by the
realization of the quantum statistics symmetry of $N$ electrons,
$S_{N}$, as the Weyl subgroup of the unitary group $U(N)$. In this
scheme, one promotes the electrons' coordinates to the eigenvalues
of two $N \times N$ hermitian matrices, and declares $U(N)$ to be a
gauge symmetry, so that the number of gauge-invariant degrees of
freedom remains the same (the $N$ `electron positions'). The
Lagrangian fluid description then yields the noncommutative
generalization of (\ref{cse}), \beq\label{nccs}
 S_{\rm eff}={k
\over 4\pi} \int \left(A\wedge_\star dA + {2i\over 3} \;A
\wedge_\star A \wedge_\star A \right) \;, \eeq where the Moyal
exterior product is defined by $\alpha \wedge_\star \beta = \alpha_i
\star \beta_j \,dx^i \wedge dx^j$ and the ordinary Moyal product:
$$
x_1 \star x_2 = x_1 x_2 +i\theta /2\;,\qquad [x_1, x_2]_\star =
i\theta\;.
$$

The elementary area of nonlocality is set by the uniform particle
density of the fluid, $\theta = (2\pi \rho_e)^{-1}$, in some
equilibrium configuration. A first satisfactory consequence of this
effective discreteness of space is that the level (and thus the
inverse filling fraction) must be quantized, due to the
non-connectedness of the group of gauge transformations; notably,
this holds even for the $U(1)$ theory on $\mathbf{R}^2_{\theta}$
\cite{NairPoly:2001rt,BakLeePark:2001ze} (see also
\cite{Krajewski:1998uf}).

The noncommutative $U(1)_\star$ gauge theory at hand is naturally
defined on noncommutative deformations of standard manifolds, but
note that non-compact cases like ${\bf R}^2_\theta$ imply a
thermodynamic limit  in the number of electrons, $N\rightarrow \infty$. Various finite $N$ versions of
(\ref{nccs}) have been proposed as finite-matrix truncations,
appropriate for the description of Hall droplets
\cite{Polychronakos:2000nt,Hellerman:2001rj,Cappelli:2004xk,CappelliRodriguez:2006wa}
and variations thereof
\cite{Polychronakos:2001mi,Polychronakos:2001uw,MorariuPoly:2005pv}.
Here we consider a noncommutative torus ${\bf T}^2_\theta$ of area
$L^2$, and define for convenience the dimensionless noncommutativity
parameter $\Theta \equiv 2\pi\theta/L^2$. When this parameter is a
{\it rational} number, the resulting theory has special properties.
This is indeed our case, since $\Theta = 1/N$, the inverse of the
number of electrons.

The organization of the paper is as follows. In the rest of this
introduction, we argue that the topological order is insensitive to
a noncommutative deformation (\ref{ssec:TD_&_NC}), and show this
more precisely for tori using T-duality (\ref{ssec:T-dual}). In
section \ref{sec:GenNCFluids}, we propose a family of slightly
generalized toroidal Hall fluids, motivated by the T-duality
properties of (\ref{nccs}). The asymmetric levels in these models
must obey a consistency condition; section \ref{ssec:GlobalProp}
describes how the latter arises, and section \ref{ssec:calculation}
uses it in computing the topological order on a torus. Section
\ref{sec:discuss} proposes some microscopic interpretations.

\subsection{Wen's topological order and
noncommutativity}\label{ssec:TD_&_NC}

\noindent

\vspace{0.3cm}

The main physical input of (\ref{nccs}) is the noncommutative
generalization of the fluid gauge symmetry. In this regard, one  important
property of this generalization is a certain tension with the
phenomenological multifluid models. $U(1)_\star^{\otimes n}$ gauge
invariance of (\ref{csm}) seems to require a purely diagonal $K_{IJ}$
matrix, due to the non-commuting character of the Moyal product.
This is the case even when all the density parameters, $\theta$, are
equal to one another. Hence, if we demand polynomial interactions in
the Moyal product, any non-abelian enhancement of the
$U(1)_\star^{\otimes n}$ multifluid gauge symmetry is necessarily a
diagonal product of $U(n_I)_\star$ groups. Without loss of
generality, we can then concentrate on the case of a simple
$U(n)_\star $ gauge factor, i.e. we have the action (\ref{nccs}), where now
$A$ is a noncommutative one-form taking values on the Lie algebra
$u(n) \equiv {\rm Lie}[U(n)]$.

The second important property of (\ref{nccs}) is its apparent
independence of the noncommutative deformation parameter $\theta$.
Despite the infinite tower of nonlinear corrections embodied in the
Moyal product, the noncommutative model on ${\bf R}^2_\theta$ ends
up being classically equivalent to (\ref{cse}) under a
Seiberg--Witten map (c.f.
\cite{GrandiSilva:2000av,BarbonParedes:2001dw}). This suggests that
(\ref{nccs}) contains just the same topological information as
(\ref{cse}). It also implies that a noncommutative multifluid theory
with general $K_{IJ}$ matrix should exist, albeit with a degree of
nonlocality that cannot be simply parameterized by polynomials in
the Moyal product.  A more precise statement can be put forward for
the model defined on the noncommutative torus, ${\bf T}^2_\theta$.
In canonical quantization, choosing temporal gauge $A_0=0$, we are
left with the purely abelian `kinetic term' in (\ref{nccs}) for the
spatial components of the gauge field, plus the Gauss constraint
\beq\label{fcon}
 \pt_1 A_2 - \pt_2 A_1 + i \,[ A_1, A_2 ]_\star
=0 \;.
\eeq
 If the gauge field is defined with periodic boundary
conditions on ${\bf T}_\theta^2$, the natural ansatz for the
solution of (\ref{fcon}) is that $A_i \, dx^i$ be a constant
one-form over ${\bf T}_\theta^2$. But those are precisely
insensitive to the noncommutativity, so that we end up with a moduli
space of flat connections {\it independent} of $\theta$, i.e. we
have the standard moduli space of $U(n)$ flat connections, which can
be parameterized locally by angular variables $c^a$, $a=1, \dots,
n,$ running in the Cartan torus of $U(n)$, ${\bf T}_{\rm C}$.
Dividing by the residual Weyl group symmetry  leads to the orbifold
$${\cal M} = {{\bf T}_{\rm C} \times {\bf T}_{\rm C} \over W_n}$$ with
$W_n = S_n$ acting diagonally on the product of the two tori. We
quantize this moduli space by reducing the Chern--Simons action to
zero-modes:
\beq\label{fccc}
 S_{\rm flat} =
2\pi k \int dt \,\sum_{a =1}^n {\dot c}_1^a \,c_2^a \;.
\eeq
Interpreting the modding by the Weyl group as the statistical
constraint for $n$ identical bosons on the torus  $(c_1, c_2)$ of unit size, this model is isomorphic to
$n$ bosons in a Landau level of $k$ states. Hence, the dimension of
the Hilbert space (the topological order on the torus) is simply
given by the number of ways of distributing the $n$ bosons over $k$ `orbitals':
\beq\label{topoun}
{\cal D}_1={\rm dim} \;{\cal H} ({\bf T}^2_\theta) = {(n+k-1)! \over
n! (k-1)!}\;.
\eeq

Notice that the result does not depend on $\theta$, and differs in
general from other natural models with $n$ fluid components, such as
the Jain hierarchical scheme for filling fractions $\nu =n/k$ and $k
= 2ns+1$, $s\in{\bf Z}$. In this case, the topological order is
given by ${\cal D}_1 =k$. Hence, the result (\ref{topoun}) seems to pose a very
stringent limitation to any model that combines enhanced gauge
symmetry and noncommutativity.

\subsection{T-duality}\label{ssec:T-dual}

\noindent

\vspace{0.3cm}

The arguments just presented can be made more rigorous using
T-duality\footnote{Also known as Morita equivalence in the
mathematical literature \cite{Schwarz:1998qj}. Morita equivalence in
general relates among themselves different non-commutative spaces
(i.e. C*-algebras), along with -in nice cases like here- all the
bundle structures necessary to define an action. For non-commutative
tori, these transformations can be thought of as a remnant of the
full T-duality of a corresponding string \cite{Pioline:1999xg} or
matrix \cite{ConnesDouglasSchwarz:1997cr} theory.}: an $SL(2,{\bf
Z})$ group of discrete transformations between theories of different
values of $\Theta$ and different ranks and quantum numbers. Given
that $\Theta = 1/N$ is rational in our case, there is a T-duality
transformation that maps (\ref{nccs}) to a level ${\widetilde k}
=kN$ Chern--Simons model with gauge group $U({\tilde n})$, ${\tilde
n} = nN$, and defined on a commutative ($\theta =0$) two-torus of
size ${\widetilde L} =L/N$. If the original $U(n)_\star$ model was
defined with periodic boundary conditions on ${\bf T}_\theta^2$, the
T-dual $U(nN)$ non-abelian (but commutative) model is restricted to
$U(nN)$ bundles with first Chern class $ c_1 = n$. Since only the
diagonal $U(1)$ subgroup contributes to this Chern class, such
bundles can be viewed as $U(1) \times \left( SU(nN)/{\bf Z}_{nN}
\right)$ bundles with $n$ units of magnetic flux on the $U(1)$
factor and $n$ units of 't Hooft magnetic flux on the $SU(nN)/{\bf
Z}_{nN}$ subgroup (see \cite{Alvarez-GaumeBarbon:2001tv} for a more
detailed exposition).

Notice that the T-dual representation uses an unphysical torus
${\widetilde {\bf T}}^2$ of length ${\widetilde L}=L/N$. In
realistic situations, $N$ is the number of electrons in the fluid
and therefore macroscopic, ${\widetilde L}$ being extremely small.
The dual model must be regarded as a mathematical tool to define the
quantization of the original noncommutative theory. In specifying
path integration in the T-dual model, the gauge fields must be
restricted to some particular class of (non-trivial) bundles over
the torus. To this end, we can consider a three-dimensional manifold
${\widetilde {\bf D}}^3$ with boundary, $\pt {\widetilde {\bf D}}^3
= {\widetilde {\bf T}}^2$, over which the bundle extends trivially.
Then we can write the dual action as
\begin{equation}\label{dualaa}
    {\widetilde S}_{\rm eff} = {{\widetilde k} \over 4\pi}
    \int_{{\widetilde {\bf D}}^3 \times \bf R} {\widetilde \Tr}
    \;({\widetilde F} - {\widetilde \Phi}) \wedge ({\widetilde F} -
    {\widetilde \Phi})\;,
    \end{equation}
where ${\widetilde \Tr}$ denotes the trace in the fundamental
representation of $U({\tilde n}) = U(nN)$. In this expression,
${\widetilde \Phi}$ denotes a constant-curvature $U(1)$ field
strength, precisely subtracting the $n$ units of magnetic flux in
the diagonal $U(1)$ group. Hence, we can regard the model as defined
on $U(nN)$ bundles of the form $U(1) \times \left( SU(nN)/{\bf
Z}_{nN} \right)$, where the $U(1)$ factor has zero curvature and the
$SU(nN)/{\bf Z}_{nN}$ bundle has $n$ units of 't Hooft magnetic flux
through the two torus ${\widetilde {\bf T}}^2$. After we implement
this constraint on the topological structure of the bundles, we can
drop the constant background ${\widetilde \Phi}$ from the action
(\ref{dualaa}).

The canonical quantization of this model follows the usual
procedure, i.e. we must quantize the moduli space of flat
connections. For the particular $U(nN)$ bundles under consideration,
the moduli space of flat connections is the direct product of the
diagonal $U(1)$ contribution, times the flat connections of the
$SU({\tilde n})/{\bf Z}_{\tilde n}$ theory with $n$ units of 't
Hooft flux. The latter are the solutions of the equations
\begin{equation}\label{fc}
    {\widetilde U}_1 {\widetilde U}_2 = {\widetilde U}_2 {\widetilde U}_1 \;e^{2\pi i /
    N} \;,
    \end{equation}
where ${\widetilde U}_1, {\widetilde U}_2$ are
matrices in $SU({\tilde n})$. There is an irreducible solution of
this equation in $SU(N)$ matrices, given by the standard {\it clock}
and {\it shift} matrices $\Gamma_1, \Gamma_2$. Then, the general
solution of (\ref{fc}) is, up to conjugacy, ${\widetilde U}_i = W_i
\otimes \Gamma_i$, $i=1,2$, with $W_i$ a pair of commuting matrices
in $SU(n)$. Hence, the required moduli space of flat connections is
isomorphic to that of an $SU(n)$ bundle with periodic boundary
conditions. Adding back the diagonal $U(1)$ factor, one finds the
flat connections of an $U(n)$ gauge model, with reduced dynamics
given exactly by (\ref{fccc}) (see
\cite{Alvarez-GaumeBarbon:2001tv}). From here, (\ref{topoun})
follows as in the previous discussion.

We conclude that the independence of the topological order as a
function of $\theta$ can be readily obtained -for rational values of
$\theta$- in a rigorous definition of the theory via the T-duality
transformation to a twisted non-abelian, but commutative model.

\section{Generalized noncommutative fluids}\label{sec:GenNCFluids}

\noindent

\vspace{0.3cm}

We have seen in the previous section that the topological order is a
priori insensible to a noncommutative deformation. In this section,
we use the quantum definition of the toroidal fluid to propose a
natural generalization that behaves discontinuously at $\theta=0$.

One of the basic features of nonabelian noncommutative gauge
theories is the automatic promotion of $SU(n)$ gauge groups to full
$U(n)_\star$ gauge invariance. The Moyal commutator, $[A_1 ,
A_2]_\star$, responsible for the self-interaction of the gauge
fields, is proportional to
$$
[T^a, T^b] \,\cos(\shalf \theta p_1 p'_2) + \{T^a, T^b\} \,\sin(\shalf \theta p_1 p'_2)\;,
$$
where $T^a$ is a basis of generators of the $u(n)$ Lie algebra, and
$p, p'$ are the momenta entering the vertex.  Hence, the diagonal
$U(1)$, from the local decomposition $u(n) = u(1) \oplus su(n)$ of
the Lie algebra, remains coupled at generic values of the momenta.

Something peculiar happens when the theory is formulated on a torus
with rational noncommutativity parameter $\Theta$. On a torus we
have $p= 2\pi n/L$ and the anticommutator part yields
$$
\{T^a, T^b\} \,\sin(\pi \Theta n_1 n'_2)\;.
$$
In our case $\Theta = N^{-1}$ and the diagonal $U(1)$ does decouple
for momentum modes proportional to the number of electrons, $N$.
Therefore, there is an effective position-space torus of size
${\widetilde L} = L/N$, which we shall denote ${\widetilde {\bf
T}}^2$, on which the diagonal $U(1)$ modes decouple completely. In
the T-dual picture, this result is obvious, since the commutative
$U(nN)$ model is now defined on a torus of size $L/N$, and there is
no local coupling of the diagonal $U(1)$ subgroup and the $SU(nN)$
subgroup. This peculiar decoupling of $U(1)$ degrees of freedom in
rational theories makes it natural to generalize our definition of a
`noncommutative fluid' to encompass these cases, by allowing
different Chern--Simons couplings in the two sectors. The resulting
Lagrangian is most easily written in T-dual variables, as a
Chern--Simons model on the torus ${\widetilde {\bf T}}^2$ of size
${\widetilde L}=L/N$ with gauge fields in the Lie algebra of the
gauge group $ U(1)_{{\tilde \kappa} {\tilde n}} \times SU( {\tilde
N})_{\tilde k} $, and with $n$ units of 't Hooft flux on the
$SU({\tilde n})/{\bf Z}_{\tilde n}$ subgroup.  The subscripts on the
gauge groups indicate the level, or Chern--Simons coupling,  with
the following normalization convention: decomposing the dual gauge
field
$$
{\widetilde A} = {\widetilde a} \cdot {\bf 1}_{\tilde n} +
{\widetilde A}'\;, \qquad {\widetilde \Tr}\; \left({\widetilde
A}'\,\right) =0
$$
into $u(1)$ and $su(nN)$ parts, the action reads
 \beq\label{newf}
 {\widetilde S}_{\rm eff} = {{\tilde \kappa}{\tilde n}  \over 4\pi} \int_{{\widetilde {\bf T}}^2 \times \bf R}
  \left(\,{\widetilde a} \wedge d\, {\widetilde a}\,\right)
+ {{\tilde k} \over 4\pi} \int_{{\widetilde {\bf D}}^3 \times \bf R}
{\widetilde \Tr}\,\left({\widetilde F}' \wedge {\widetilde F}'
\right)\;. \eeq The classical $U(n)_\star$ model written in T-dual
variables, eq. (\ref{dualaa}), is obtained in the case ${\tilde
\kappa} = {\tilde k}$. Recall that the T-duality map implies
${\tilde n} = nN$ and ${\tilde k} = kN$. Hence, we will find it
useful to parametrize the abelian level as ${\tilde \kappa} = \kappa
N$.

Going back to the original
variables, in terms of Moyal products, we have
\beq\label{origv}
 S_{\rm
eff} = (\kappa -k) {nN^2 \over 4\pi} \int_{{\widetilde {\bf T}}^2
\times \bf R} {\widetilde a} \wedge d\,{\widetilde a} + {k \over
4\pi} \int_{{\bf T}^2_\theta \times \bf R} \Tr \left( A\wedge_\star
d\,A + {2i\over 3} A\wedge_\star A \wedge_\star A \right) \;.
\eeq
 In this expression we have kept the dual notation for the
purely abelian term, in order to have a local expression in position
space. In the original representation on ${\bf T}^2_\theta$, we must
specify to keep only those diagonal $U(1)$ degrees of freedom whose
momentum quantum numbers vanish modulo $N$.

The asymmetry between the abelian and non-abelian sectors is
measured by the difference $\kappa -k$. For $\kappa =k$, we recover
the lagrangian (\ref{dualaa}) of the previous section. However,
different choices of $\kappa$ might be natural depending on
particular circumstances. For example, if the original
noncommutative gauge field couples to {\it massive} degrees of
freedom in the adjoint representation, integrating them out induces
a shift of the low-energy Chern--Simons level at one loop order.
Thus for each bosonic degree of freedom in the adjoint
representation the level shifts $k\rightarrow k+n$, whereas the
shift is only half as much and negative, $k\rightarrow k - \shalf n$
for each real fermionic (Majorana) degree of freedom in the adjoint
representation. On the rational noncommutative torus, all these
shifts leave intact the diagonal $U(1)$ fields with momenta
proportional to $N$ (see \cite{Vassilevich:2007gt} for a recent
explicit check). Hence, by integrating out massive adjoints we can
end up with an effective $\kappa$ that differs from $k$.

On the other hand, Wilson loop expectation values in an ordinary $SU(n)$
Chern--Simons model are naturally a function of $k+n$, whereas the
abelian counterparts lead to simple functions of $k$. In order to have a
simple $U(n)$ symmetry action over quantum Wilson lines, it is
useful to start with a classical model of the form (\ref{origv}) with
$\kappa = k+n$. Hence, we see various instances in which
consideration of the generalized fluids (\ref{origv}) might be regarded as
natural.

\subsection{Global properties}\label{ssec:GlobalProp}

\noindent

\vspace{0.3cm}

What we have said so far only refers to couplings in the Lagrangian.
However, the precise definition of the non-abelian fluid models
requires further discrete projections, particularly if we wish to
recover the $U(n)_\star$ model at $\kappa =k$, discussed in the
previous section. The global structure of the $U(n)$ group, i.e. $
U(n) = \left(U(1) \times SU(n)\right) / {\bf Z}_n$ implies that the
flat connections in the diagonal $U(1)$ are correlated with those of
the non-abelian factor by the modding by the centre of $SU(n)$. That
is, the ${\bf Z}_n$ redundancy $U= u\, U' = u \,z \cdot z^{-1}
\,U'$, with $u\in U(1)$, $U' \in SU(n)$ and $z \in {\bf Z}_n$, in
the decomposition of an arbitrary $U(n)$ matrix, is treated as a
discrete gauge symmetry and the theory is projected onto the ${\bf
Z}_n$-invariant sector.

More specifically, the ${\bf Z}_n$ group acts by `large' gauge
transformations on the $SU({\tilde n})/{\bf Z}_{\tilde n}$ factor.
For Chern--Simons theories, it is enough to specify this action at
the level of flat connections on the torus. On a bundle with $n$
units of 't Hooft flux, flat connections are specified by the
holonomies ${\widetilde U}_1, {\widetilde U}_2$ in the two
directions of the torus. As pointed out above, these have the form
${\widetilde U}_i = W_i \otimes \Gamma_i$ with $W_i$ a pair of
commuting matrices in $SU(n)$ and $\Gamma_i$ the (essentially
unique) solution of $\Gamma_1 \Gamma_2 = \Gamma_2 \Gamma_1
\,\exp(2\pi i /N)$ in $SU(N)$ matrices. Then, there is a ${\bf Z}_n
\times {\bf Z}_n$ group of large gauge transformations acting as
$$
W_j \rightarrow z_j \,W_j\;, \;\;\;\;j=1,2\;,
$$
with $z_j = \exp(2\pi i \alpha_j /n)$, $\alpha_j \in {\bf Z}$. On
the Cartan torus of the $W_i$, the group ${\bf Z}_n$ acts by
discrete translations proportional to $\alpha_j /n$. The flat
holonomies on the abelian diagonal $U(1)$ are given by a pair of
phases $w_j $, acted upon by the ${\bf Z}_n \times {\bf Z}_n$ group
as
$$
w_j \rightarrow z_j^{-1}\, w_j \;,\;\;\;\;j=1,2\;,
$$
so that the total holonomy $w_j W_j$ remains ${\bf Z}_n$-invariant.

Upon canonical quantization, wave functionals can be chosen in
irreducible representations labelled by the associated characters,
which are known as 't Hooft electric flux sectors in ordinary gauge
theories. In topological Chern--Simons theories, the flat holonomies
along each direction of the torus end up as canonical conjugates of
one another. Therefore, only one set of holonomies, and their
corresponding  set of electric flux sectors, is used to label
physical states. Following this procedure, let us conventionally
choose $w_1 \in U(1)$, $ W_1 \in SU(n)$ as the flat holonomies
parametrizing the configuration space of the system. The Hilbert
space of the level-$\kappa n$ abelian Chern--Simons theory consists
of wave functions $\psi_{e_1} (w_1)$, partially classified by their
representation under ${\bf Z}_n$, i.e. by the value of the electric
flux $e_1$, an integer defined modulo $n$ by the transformation:
$$
\psi_{e_1} (w_1) \rightarrow \psi_{e_1} \left(z_1^{-1} w_1\right) =
(z_1)^{-e_1}\; \psi_{e_1} (w_1)
$$
For the non-abelian $SU({\tilde n})/{\bf Z}_{\tilde n}$ sector we
have analogous electric flux sectors $e'_1$,
$$
\Psi_{e'_1} \left(W_1 \right) \rightarrow \Psi_{e'_1} \left( z_1
W_1\right) = (z_1)^{e'_1} \;\Psi_{e'_1} \left( W_1 \right)\;,
$$
also given as integers modulo $n$. When collecting  together the
abelian and non-abelian sectors, the projection to the singlet
representation of ${\bf Z}_n$ imposes  the constraint
$$
e_1 = e'_1 \;\;({\rm mod}\;\;\;n)
$$
which effectively reduces the dimension of the Hilbert space by a
factor of $n$ with respect to the dimension of the product theory
$U(1)_{\kappa n} \times SU(n)_k$. An analogous projection must be
enforced for the conjugate `momentum' variables, $w_2, W_2$. If the
latter is consistent with the first projection, we should end up
with a reduction of the naive product dimension by a factor of
$n^2$. In the next section we confirm this counting by a more
detailed analysis of the ${\bf Z}_n$ action in a convenient basis.

The total dimension of the $SU(n)$ Chern--Simons Hilbert space is
given by the number of integrable representations of the $SU(n)_k$
Kac--Moody algebra. This number can be computed by counting $SU(n)$
Young tableaux with at most $k$ boxes, yielding
$$
{\cal D}_1 (SU(n)_k) = {(n+k-1)! \over (n-1)! \,k!}
$$
On the other hand, the level-$\kappa n$ abelian Chern--Simons model
has $\kappa n$ states. Hence, we are predicting \beq\label{pred}
{\cal D}_1 (\kappa, n, k) = {\kappa (n+k-1)! \over n! \,k!} \eeq as
conjectured in \cite{Alvarez-GaumeBarbon:2001tv}. The condition that
this generalized topological order be an integer requires that
$\kappa = k \; ({\rm mod} \;n)$. In the next section we flesh out
this constraint as a consistency condition in the quantization
procedure.

\subsection{Calculation of the generalized topological
order}\label{ssec:calculation}

\noindent

\vspace{0.3cm}

In this section we compute the topological order of the generalized
fluids, confirming the conjecture (\ref{pred}). As explained in
previous sections, it suffices to quantize the space of naive zero
modes of a Chern--Simons model
$$
{U(1)_{\kappa n} \times SU(n)_k \over {\bf Z}_n}
$$
where the group  ${\bf Z}_n$ is interpreted as a discrete gauge
symmetry. The reduced configuration space to be quantized consists
of the flat holonomies, with associated  connections parameterizing
the moduli space \beq\label{redm} {{\bf T}_{\rm C} \times {\bf
T}_{\rm C} \over ({\bf Z}_n \times {\bf Z}_n)\times W_n}\,. \eeq
 As before, the two $n$-tori in the numerator arise from the
two cycles of the spatial torus. We parameterize these holonomies by
diagonal elements $C^0$ and ${\bf C}$ in $u(n)=u(1)\oplus su(n)$,
with the identifications
$$C^0_i\sim C^0_i + 1\,,\,\;\;{\rm and} \qquad {\bf C}_i\sim {\bf C}_i+{\bf
\alpha}\,,\qquad (i=1,2)$$ for any (co-)root ${\bf \alpha}\in
\Lambda_R$.

The residual Weyl symmetry comes from constant gauge transformations
$g(x)=g_{\Gamma}\in SU(n)$ acting by conjugation on both holonomies
simultaneously:
\beq\label{weyla}
 W_n \ni \Gamma:\quad \left({\bf
C}_1,{\bf C}_2\right) \longrightarrow \left(\Gamma^{-1}({\bf
C}_1),\Gamma^{-1}({\bf C}_2)\right)\,.
\eeq
 The division by ${\bf Z}_n$
accounts for the freedom in lifting a path from $U(n)$ to
$U(1)\times SU(n)$. A generator of this group acts, on each
$n$-torus separately, by a translation \beq\label{zna}
 {\bf Z}_n \ni {\bf
T}^{(j)}_i:\quad \left(C^0_i,{\bf C}_i\right) \longrightarrow
\left(C_i^0 - {j \over n},{\bf C}_i - {\bf w}_j\right)\,\qquad
i=1,2
\eeq
 for some integer $j$ relatively prime to $n$. For
definiteness, we use the trace in the fundamental representation
$<\cdot,\cdot>\equiv Tr(\cdot,\cdot)$ to identify $u(n)$ with its
dual. The fundamental weights ${\bf w}_j,j=1,..,n-1$ are defined as
the basis of $su(n)$ dual to the roots $\alpha_j\equiv
diag(0,..,1_j,-1_{j+1},..,0)$. Note that $r{\bf w}_j\in\Lambda_R$
only for $r$ a multiple of $n$.

In fact the levels play the role of two scaling factors, for the
scalar product on $u(1)$ and for the Killing form. It will be
convenient to decompose ${\bf C}_1=C_1^a {\bf \alpha}_a$ in terms of
roots, and ${\bf C}_2=C_2^a {\bf w}_a$ in terms of weights, to have
simply $\left<{\dot {\bf C}}_1,{\bf C}_2\right> = \sum_{a=1}^{n-1}
{\dot C}_1^a C_2^a\,.$ The reduced action
\beq\label{reda}
 S_{\rm flat} =
2\pi \kappa n \int dt \;{\dot C}^0_1\, C^0_2 + 2\pi k \int dt\;
\sum_{a=1}^{n-1} {\dot C}_1^a C_2^a\,.
\eeq
 gives canonical commutation relations
$$\left[C^0_1,C^0_2\right]={i \over 2\pi \kappa n}\quad
\left[C^a_1,C^b_2\right]={i\delta^{ab} \over 2\pi k}$$ which fit
nicely with the product structure in the numerator of the orbifold
(\ref{redm}) \footnote{We shall use a `bosonic' representation for
the wave functions, in which the Weyl group acts as a statistical
group on an effective bosonic Hilbert space. A more standard basis
used in the literature, which duly incorporates the quantum shift $k
\rightarrow k+n$ of the non-abelian theory, involves a `fermionic'
action of the Weyl group (c.f.
\cite{ElitzurMooreSchwimmerSeiberg:1989nr,Douglas:1994ex,LabastidaLlatasRamallo:1990bt}.
For the purposes of this note, the bosonic representation as used in
\cite{WenZee:1997ce}\cite{Witten:1999ds}, is actually more
convenient.}. It allows to construct the Hilbert space starting with
functions in one set of $n$ variables, say $C^0_1,C^a_1$, which may
be thought of as "positions", and represent in the standard way the
conjugate variables, i.e. the `momenta' $C^0_2,C^a_2$, as derivative
operators. These wave functions are linear superpositions,
$\Psi(C^0_1,{\bf C}_1)=\sum_{\left(m,{\bf M}\right)\in {\bf Z}\times
\Lambda_w} \Psi_{m,{\bf M}}\, \chi_{m,{\bf M}}$, of characters
\begin{equation}\label{chia} \chi_{m,{\bf M}}\equiv \exp{2\pi
i\left(mC_1^0+\left< {\bf M},{\bf C}_1 \right>\right)}
\end{equation}
labeled by the abelian and non-abelian weights $m$ and ${\bf M}$.
Representing the c.c.r. on this space leads to the identifications
$m\sim \kappa n C^0_2$ and ${\bf M}\sim k{\bf C_2}$. First, this
furnishes an alternative interpretation for the characters, as delta
functions in the conjugate variables \beq\label{chib} \chi_{m,{\bf
M}}(C^a_2) = \delta (\kappa n C^0_2 - m)\, \delta(k{\bf C}_2 - {\bf
M}). \eeq Second, the total Hilbert space ${\hat {\cal H}}$ -i.e.
the result of quantizing ${{\bf T}_{\rm C} \times {\bf T}_{\rm C} /
({\bf Z}_n \times {\bf Z}_n)}$- is effectively finite dimensional
and consists of equivalence classes of wave functions labeled by
$\left(m,{\bf M}\right)\in {{\bf Z}_{\kappa n}}\times \Lambda_w /
k\Lambda_R$. Now, the modding of ${\hat {\cal H}}$ by the Weyl group
and ${\bf Z}_n$ is to be obtained from their actions (\ref{weyla}
and \ref{zna}) on the underlying phase space (\ref{redm}). An
element $\Gamma\in {\bf W}_n$ sends $\chi_{m,{\bf M}}$ to
$\chi_{m,\Gamma({\bf M})}$, consistently in both representations
(\ref{chia}) and (\ref{chib}) . Similarly, one easily deduces the
effect of ${\bf Z}_n$, for holonomies number one and number two. The
relevant actions on wave functions are
\begin{eqnarray*}
    \Gamma:& \Psi_{m,{\bf M}} \longrightarrow & \Psi_{m,\Gamma^{-1}\left({\bf
    M}\right)}\\
    T^{(j)}_1:& \Psi_{m,{\bf M}} \longrightarrow & \Psi_{m,{\bf M}}\,
    \exp{2\pi i\left(-{mj \over n}-<{\bf M},{\bf w}_j>\right)} \\
    T^{(j)}_2:& \Psi_{m,{\bf M}} \longrightarrow &
    \Psi_{m-j\kappa,{\bf M}-k{\bf w}_j}.
    \end{eqnarray*}
Computing their commutators gives
\begin{eqnarray*}
    \left[T_1^{(j)}\Gamma\,(T_1^{(j)})^{-1}\Gamma^{-1}\right]\Psi_{m,{\bf M}}&=&
    \Psi_{m,{\bf M}}\,\exp{2\pi i<\Gamma({\bf M})-{\bf M},{\bf
    w}_j>}\\
    &=& \Psi_{m,{\bf M}}\\
    \left[T_2^{(j)}\Gamma\,(T_2^{(j)})^{-1}\Gamma^{-1}\right]
    \Psi_{m,{\bf M}}&=&\Psi_{m,{\bf M}+k{\bf w}_j-\Gamma(k{\bf
    w}_j)}\\
    &\sim & \Psi_{m,{\bf M}}
    \end{eqnarray*}
using the fact that $\Gamma({\bf w})-{\bf w}\in \Lambda_R$ for any
${\bf w}\in \Lambda_w$, and
$$[T_1^{(j)}T_2^{(l)}(T_1^{(j)})^{-1}(T_2^{(l)})^{-1}]
    = \exp{-2\pi i \,\left(\kappa {j l\over n}+<k{\bf w}_l,{\bf w}_j>\right)}
    = \exp{2\pi i \,{jl\over n}\,(k-\kappa)}$$
as the scalar products of the fundamental weights are
$\left<w_j,w_l\right> = min(j,l)-jl/n$ (the inverse of the Cartan
matrix).

Thus, we find that, in order to obtain a non-trivial quantum theory
(i.e. $dim({\cal H})\neq 0$), the abelian and non-abelian levels must
obey the compatibility condition
$$
\kappa = k + s n
$$
with $s\in {\bf Z}$.

In this case, the modding by the denominator in (\ref{redm}) amounts in
fact, as far as its effect on ${\hat {\cal H}}$ is concerned, to a
faithful action of ${\bf Z}_n \times {\bf Z}_n \times W_n$.
In particular, the topological order of the general model can be
written as $$\CD_1(\kappa,k,n) = {\CD_1(U(1)_{\kappa n})\cdot
\CD_1(SU(n)_k) \over n^2},$$ and using for calibration its value in
the known cases of $U(1)$ and $U(n)$, one has
$$
\CD_1(\kappa,k,n) = {\kappa \cdot (n+k-1)! \over n! k!}= \CD_1
(U(n)_k) + s \CD_1 (SU(n)_k).
$$
Incidentally, this relation also provides an easy way to compute
$\CD_1 (SU(n)_k)$.

\section{Discussion}\label{sec:discuss}

\noindent

Noncommutative Chern--Simons models for Hall fluids show little or
no dependence on the deformation parameter $\theta$. In particular,
the quantum structure of the theory on a compact two-torus is
explicitly independent of the rational noncommutativity parameter
$\Theta$, a conclusion that follows from careful consideration of
the T-duality symmetry.  In this note, we have argued that the
assumption of a noncommutative effective action does impose some
constraints on the universality classes of {\it nonabelian} fluids,
in the form of discrete selection rules.

At the classical level, the statistical gauge symmetry is restricted
to be a diagonal product of $U(n)_\star$ gauge groups. In
particular, multifluid models of the form (\ref{csm}) with
non-diagonal $K$ matrix do not admit a polynomial action in Moyal
products. Instead, any non-abelian statistics group with an $SU(n)$
factor is forced to combine with a diagonal  $U(1)$ into a full
$U(n)_\star$ gauge group. At the quantum level, the rationality of
the noncommutative theory on the torus allows a certain relaxation
of this rule, selecting universality classes characterized by groups
$\left[U(1)_{\kappa n} \times SU(n) \right]/{\bf Z}_n$, where the
abelian level is restricted by the selection rule $\kappa = k
\;({\rm mod}\;n)$.

Among the proposed universality classes, the case $\kappa = k+n$ is
especially interesting. For this choice of abelian level,
correlation functions of Wilson-line operators have a uniform
dependence on the combination $k+n$, suggesting a quantum
realization of the $U(n)$ gauge symmetry (c.f.
\cite{MarinoVafa:2001re}). The associated WZW model of edge currents
is also the standard definition of the $u(n)$ Kac--Moody algebra
\cite{NaculichSchnitzer:2007nc}. Finally, this case yields a natural
generalization of the `parton construction' of Hall fluids with
nonabelian statistics in \cite{WenZee:1997ce}, to which the
concluding remarks are devoted.

Following \cite{WenZee:1997ce}, we can heuristically construct
hydrodynamical models with non-abelian statistics by splitting the
electron into partons, that are subsequently bound together into
full electrons by means of a non-abelian gauge field. The
construction presented explicitly in \cite{WenZee:1997ce} refers in
our notation to the case $n=2$, generating fluids with topological
order $(k+1)(k+2)/2$ and wave function ${\cal X}_1 ({\cal X}_k)^2$, with
${\cal X}_k$ the standard wave function for $k$ filled Landau levels.

The generalization to fluids with wave function ${\cal X}_1 ({\cal X}_k)^n$
and $U(n)$ quantum statistics proceeds as follows. We split the
electron into $n+1$ partons. The first $n$ partons, with fractional
electric charge $q_s = 1/(k+n)$, are bound by an $SU(n)$ gauge field
into a baryon-like configuration, which is then bound to the
remaining parton by the diagonal $U(1)$ gauge group. The
second-quantized formalism  at the parton level includes field
operators $\Phi_{(k)}$ in the fundamental representation of $SU(n)$
and filling $k$ Landau levels, together with field operators
$\phi_{(1)}$ with unit Landau multiplicity and charged only under
the diagonal $U(1)$ group (since the partons are regarded as fermions, we require $n$ to be an even integer, for the electron bound state to remain a fermion). The electric charge of the remaining
parton $q_d = k/(k+n)$ ensures that $q_d + n q_s =1$, the electron
charge.  The resulting Hamiltonian reads
$$
H_{\rm parton} = {1\over 2\mu} \phi^\dagger_{(1)} \left( -i{\bf \nabla} -
q_d {\bf A}_{\rm em} +\Tr {\bf A} \right)^2 \,\phi_{(1)} + {1\over
2\mu} \Phi_{(k)}^\dagger \,\left( -i{\bf \nabla} - q_s {\bf A}_{\rm em} -
{\bf A} \,\right)^2 \,\Phi_{(k)}
$$
In this expression, ${\bf A}$ denotes the $U(n)$ gauge field,
$\Tr {\bf A} /n$ is the diagonal $U(1)$ gauge field, and ${\bf A}_{\rm em}$ is the
external electromagnetic field. Notice that the $n$ partons with
charge $q_s$ transform in the fundamental representation of the
statistical $U(n)$ group. The baryonic states in the totally antisymmetric
product of $n$ fundamentals are  $SU(n)$ invariant, but of course
transform with charge $n$ with respect to the diagonal $U(1)$
subgroup. This is the reason for the introduction of the extra
parton, a singlet of $SU(n)$, and canceling the $U(1)$ charge of the
`baryon'. Finally, since the partons are treated as fermions, $n$ must be restricted to even values to ensure the correct physical interpretation of the electron bound state.

Upon integrating out the parton fields $\phi, \Phi$, as in  \cite{WenZee:1997ce},
 we generate
various Chern--Simons actions. A purely electromagnetic
Chern--Simons action with  conductivity coefficient
$$
\nu= q_d^2 + k n q_s^2
\;,$$
a level-$k$ statistical Chern--Simons action for  the $SU(n)$  field ${\bf A} - \Tr {\bf A} /n$, and an abelian
Chern--Simons action on the field $\Tr{\bf A}/n$ with level
$$
\kappa n = n^2 + n k
$$
For our particular assignment of electric charges,  $q_s = 1/(k+n),
q_d = k/(k+n)$, we obtain a family of  hydrodynamical phases with
filling fraction $\nu = k/(k+n)$, realizing our noncommutative
universality classes with $\kappa = k+n$. As we can see from the
explicit form of the Hamiltonian, this particular linking of the
abelian and non-abelian levels is borne out by imposing that the
`baryonic' partons transform in the fundamental representation of
the full $U(n)$ statistical group. We can also generate all  other
models with $\kappa = k+ s n$ by imposing irrational $U(1)$ charges
with respect to the statistical field $\Tr {\bf A}$. This is much less
satisfactory, since we lose the natural action of a $U(n)$ symmetry
at the level of the quasiparticle, `parton'  Hamiltonian.

\vskip 1cm

{\bf Acknowledgements}

\noindent

 This work was partially supported by MCyT
 and FEDER under grant FPA2006-05485, CAM under grant HEPHACOS P-ESP-00346
 and the European RTN network
 MRTN-CT-2004-005104.


\bibliographystyle{utphys}
\bibliography{references}
\end{document}